\documentclass[pra,a4paper,superscriptaddress,twocolumn,showpacs,amsmath,amssymb,floatfix]{revtex4}

\usepackage[dvips]{graphicx}
\usepackage{bm,pstricks,longtable}

\newcommand{\ket}[1]{| #1\rangle}                       %

\newcommand{\bra}[1]{\langle #1}                        %

\def\eps{\varepsilon}
\begin{document}

\title{Diffusion and localization for the Chirikov typical map}

\author{Klaus M. Frahm}
\affiliation{\mbox{Universit\'e de Toulouse, UPS,
Laboratoire de Physique Th\'eorique (IRSAMC), F-31062 Toulouse, France}}
\affiliation{\mbox{CNRS, LPT (IRSAMC), F-31062 Toulouse, France}}
\author{Dima L. Shepelyansky}
\affiliation{\mbox{Universit\'e de Toulouse, UPS,
Laboratoire de Physique Th\'eorique (IRSAMC), F-31062 Toulouse, France}}
\affiliation{\mbox{CNRS, LPT (IRSAMC), F-31062 Toulouse, France}}

\date{appeared in Phys. Rev. E 80, 016210 (2009)}

\begin{abstract}
We consider the classical and quantum properties of the "Chirikov typical
map", proposed by Boris Chirikov in 1969. This map is obtained from 
the well known Chirikov standard map by introducing a finite number 
$T$ of random 
phase shift angles. These angles induce a random behavior for small time 
scales ($t<T$) and a $T$-periodic iterated map which is relevant for larger 
time scales ($t>T$). We identify the classical chaos border 
$k_c\sim T^{-3/2} \ll 1$ for the kick parameter $k$ and two regimes with 
diffusive behavior on short and long time scales.
The quantum dynamics is characterized by the effect of 
Chirikov localization (or dynamical localization).
We find that the localization length depends in a subtle way on the two 
classical diffusion constants in the two time-scale regime.

\end{abstract}


\maketitle

\section{Introduction}
\label{sec1}
The dynamical chaos in Hamiltonian systems often leads to
a relatively rapid phase mixing and a relatively slow 
diffusive spreading of particle density
in an action space \cite{lichtenberg}. A well known
example of such a behavior is given by the Chirikov standard map
\cite{chirikov1,chirikov2}.
This simple area-preserving map appears in a description
of dynamics of various physical systems
showing also a generic behavior of chaotic Hamilton systems
\cite{scholar}. The quantum version of this 
map, known as the quantum Chirikov standard map or kicked rotator, 
shows a phenomenon of 
quantum localization of dynamical chaos
which we will call the Chirikov localization,
it is also known as the dynamical localization.
This phenomenon had been first seen in numerical simulations
\cite{kr1979} while the dependence of the localization
length on the classical diffusion rate and the Planck
constant was established in 
\cite{chi1981,chidls,dls1986}.
It was shown in \cite{prange} that this localization is
analogous to the Anderson localization of quantum waves in
a random potential (see \cite{fishman} for more Refs. and details).
In this respect the Chirikov localization can be viewed as a dynamical
version of the Anderson localization: in dynamical
systems diffusion appears due to dynamical chaos while 
in a random potential diffusion appears due to disorder
but in both cases the quantum interference leads to localization
of this diffusion.  The quantum Chirikov standard
map has been built up experimentally with cold atoms in kicked
optical lattices \cite{raizen}. In one dimension all
states remain localized.
In systems with higher dimension
(e.g. $d=3$) a transition from localized to delocalized 
behavior takes place as it has been demonstrated in
recent experiments with cold atoms in kicked optical lattices
\cite{garreau}.

While the Chirikov standard map finds various applications
it still corresponds to a regime of kicked systems
which are not necessarily able to describe a continue
flow behavior in time. To describe the properties of such a 
chaotic flow Chirikov introduced in 1969 \cite{chirikov1}
a typical map which we will call the Chirikov
typical map. It is obtained by $T$ iterations of 
the Chirikov standard map with random phases which 
are repeated after $T$ iterations. For small kick amplitude this model
describes a continuous flow in time with the Kolmogorov-Sinai
entropy being independent of $T$. In this way this model
is well suited for description of chaotic continuous 
flow systems, e.g. dynamics of a particle in random magnetic fields
\cite{rechester} or ray dynamics in rough billiards \cite{frahm1,frahm1a}.
Till present only certain  properties of the classical typical
map  have been considered in  \cite{chirikov1,chi1981}.

In this work we study in detail the classical diffusion and 
quantum localization in the Chirikov typical map. Our results
confirm the estimates presented by Chirikov \cite{chirikov1}
for the classical diffusion $D_0$ and instability. For the quantum model
we find that the localization length $\ell$ is determined by the
classical diffusion rate per a period of the map $\ell \sim D_0T/\hbar^2$ 
in agreement with the theory \cite{chidls,dls1986}.

The paper is constructed in the following way:
Section II gives the model description, dynamics properties and chaos border
are described in Section III,
the properties of  classical diffusion
are described in Section IV, 
the Lyapunov exponent and instability properties are considered
in Section V,
the quantum evolution is analyzed in Section VI,
the Chirikov localization is studied in Section VII and the discussion
is presented in Section VIII.

\section{Model}
\label{sec2}
\begin{figure}[ht]
\begin{center}
\includegraphics[width=0.35\textwidth]{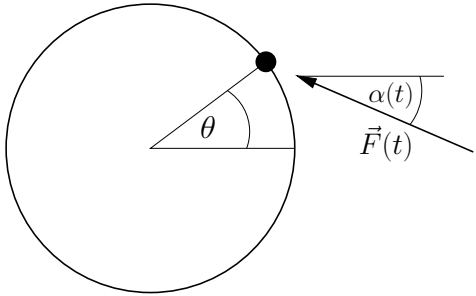} 
\end{center}
\caption{
Kicked rotator where the kick force rotates in time with a
kick force  angle 
$\alpha(t)$ that is a periodic function in time.
\label{fig1}}
\end{figure}

The physical model describing the Chirikov typical map can be
obtained in a following way. 
We consider a kicked rotator with a kick force rotating in time (see Fig. 
\ref{fig1}):
\begin{equation}
\label{kickforce}
\vec F(t)=k\left(\begin{array}{r}
-\cos \alpha(t)  \\
 \sin \alpha(t) \\
\end{array}\right)\,\sum_{n=-\infty}^\infty \delta(t-n)
\end{equation}
where $k$ is a parameter characterizing the amplitude of the kick force. 
We measure the time in units of the elementary kick period and we 
assume that the angle $\alpha(t)$ is a $T$-periodic function 
$\alpha(t+T)=\alpha(T)$ with $T$ being integer (i.~e. an integer multiple of 
the elementary kick period). The time dependent Hamiltonian associated to 
this type of kicked rotator reads:
\begin{equation}
\label{rotorham}
H(t)=\frac{p^2}{2}+k\cos[\theta+\alpha(t)]\,\sum_{n=-\infty}^\infty 
\delta(t-n)\ .
\end{equation}

To study the classical dynamics of this Hamiltonian 
we consider the values of $p$ and 
$\theta$ slightly before the kick times: 
\begin{displaymath}
p_t=\lim_{\eps\searrow 0} p(t-\eps)\quad,\quad
\theta_t=\lim_{\eps\searrow 0} \theta(t-\eps)
\end{displaymath}
where $t$ is the integer time variable. The time evolution is 
governed by the Chirikov typical map defined as:
\begin{equation}
\label{classicalmap}
p_{t+1}=p_t+k\,\sin(\theta_t+\alpha_t)\quad,\quad 
\theta_{t+1}=\theta_t+p_{t+1}
\end{equation}
with $\alpha_t=\alpha(t)$ at integer times. Since 
$\alpha_t=\alpha_{t \mod T}$ there are $T$ independent different 
phase shifts $\alpha_t$ for $t\in\{0,\, \ldots,\,T-1\}$. 
Following the approach of Chirikov \cite{chirikov1}
we assume that these $T$ phase shifts are independent and 
uniformly distributed 
in the interval $[0,2\pi[$. 

For the quantum dynamics we consider the quantum state slightly before 
the kick times:
\begin{displaymath}
\ket{\psi_t}=\lim_{\eps\searrow 0}\ket{\psi(t-\eps)}
\end{displaymath}
whose time evolution is governed by the quantum map:
\begin{equation}
\label{quantummap}
\ket{\psi_{t+1}}=\exp\left(-i\frac{\hat p^2}{2\hbar}\right)
\,\exp\left(-i\frac{k}{\hbar}\,\cos(\hat\theta+\alpha_t)\right)\,
\ket{\psi_t}\quad ,
\end{equation}
with $\hat p = -i\hbar \partial/\partial \theta$ and the wave function
$\psi(\theta+2\pi)=\psi(\theta) = \langle \theta | \psi \rangle$.
In the following we refer to the map (\ref{classicalmap}), 
(\ref{quantummap}) as the classical or quantum version of the 
{\em Chirkov typical map} as it was introduced by Chirikov \cite{chirikov1}. 
The Chirikov standard map \cite{chirikov2}
corresponds to a particular choice of random 
phases all being equal to a constant. 

\section{Classical dynamics and chaos border}
\label{sec3}

It is well known that the Chirikov standard map exhibits a transition to 
global (diffusive) chaos at $k>k_c\approx 0.9716$ 
\cite{lichtenberg}. For the Chirikov typical map 
(\ref{classicalmap}) it is possible to observe the transition to global chaos 
at values $k_c\ll 1$ provided that $T\gg 1$. In order to 
determine $k_c$ quantitatively we apply 
the Chirikov criterion of  overlapping resonances \cite{resonances}
(see also \cite{chirikov1,chirikov2} for more details). For this 
we develop the $T$-periodic kick potential in (\ref{rotorham}) in 
the Fourier series:
\begin{equation}
\label{hamfourier}
H(t)=\frac{p^2}{2}+\Re\left(\sum_{m=-\infty}^\infty 
f_m e^{i(\theta-m\omega t)}\right)
\end{equation}
with:
\begin{equation}
\label{fmdef}
\omega=\frac{2\pi}{T}\quad,\quad f_m=\frac{k}{T}\sum_{\nu=0}^{T-1}
e^{im\omega \nu+i\alpha_\nu}\quad.
\end{equation}
The resonances correspond to $\theta(t)=m\omega t$ with integer $m$ and 
positions in $p$-space: $p_{\rm res}=m\omega=2\pi m/T$. Therefore the distance 
between two neighboring resonances is $\Delta p=\omega=2\pi/T$ and the 
separatrix width of a resonance is
$4\sqrt{|f_m|}\approx 4 \langle |f_m|^2\rangle^{1/4}
=4\left({k^2}/{T}\right)^{1/4} $ where the average is 
done with respect to $\alpha_\nu$. The transition 
to global chaos with diffusion in $p$ takes place when the resonances 
overlap, i.~e. if their width is larger than their distance 
which corresponds to the chaos border
\begin{equation}
\label{kctyp}
k > k_c = \pi^2/(4 T^{3/2})
\end{equation} 
with $k_c\ll 1$ for $T\gg 1$.

\begin{figure}[ht]
\begin{center}
\includegraphics[width=0.48\textwidth]{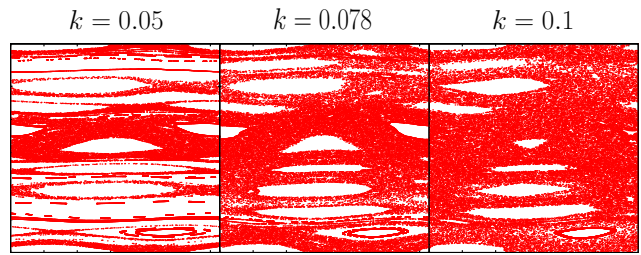} 
\end{center}
\caption{
Classical Poincar\'e sections of 50 trajectories of length 
$t_{\rm max}=10000$ of the Chirikov typical map for one particular 
realization of the random phase shifts with $T=10$ and three 
values: $k=0.05$, $k=0.078$, $k=0.1$. The positions 
$(\theta_t,p_t)\in [0,2\pi[ \times [0,2\pi[$ are shown 
after applications of the 
full map, i.~e. $T$-times iterated typical map with: $t \mod T=0$.
The value of the middle figure corresponds 
to the critical value $k_c \approx 0.078$ of (\ref{kctyp}) for the 
transition to global diffusive chaos. 
\label{fig2}}
\end{figure}

It is well known \cite{lichtenberg,chirikov1,chirikov2} 
that the Chirikov criterion of overlapping resonances 
gives for the Chirikov standard map a numerical value 
$\pi^2/4\approx 2.47$ which is
larger than the real critical value $\approx 0.9716$ that is due 
to a combination of various reasons (effect of second order resonances, 
finite width of the chaotic layer etc.). However, for the Chirikov 
typical map 
these effects appear to be less important, probably due of the 
random phases in the amplitudes $f_m$ and phase shifts of resonance positions.
Thus, the expression 
(\ref{kctyp}) for the chaos 
border works quite well including the numerical prefactor, 
even though the exact value $k_c$ depends also on 
a particular realization of random phases. 
In Fig. \ref{fig2}
we show the Poincar\'e sections of the Chirikov typical map for a particular 
random phase realization for $T=10$ and three different values of 
$k\in\{0.05, 0.078, 0.1\}$ where the critical value at $T=10$ is 
$k_c=(\frac{\pi}{2})^2\,10^{-3/2}\approx 0.078$. Fig.~\ref{fig2} confirms 
quite well that the transition to global chaos happens at that value. 
Actually choosing one particular initial condition 
$\theta_0=0.8\cdot 2\pi$ and $p_0=0.25\cdot 2\pi$ one clearly sees that 
at $k=0.05$ only one invariant curve at $p\approx p_0$ is filled and 
that at $k=0.1$ nearly the full elementary cell is filled diffusively. 
At the critical value $k=0.078$ only a part of the region 
$p>p_0$ is diffusively filled during a given number of map iterations.

\section{Classical diffusion}
\label{sec4}

For $k>k_c$ the classical dynamics becomes diffusive in $p_t$ and for 
$k\gg k_c$ we can easily evaluate the diffusion constant assuming that 
the angles $\theta_t$ are completely random and uncorrelated. In order to 
discuss this in more detail we iterate the classical map (\ref{classicalmap}) 
up to times $t$:
\begin{equation}
\label{map_p_iter}
p_t=p_0+k\sum_{\nu=0}^{t-1} \sin(\theta_\nu+\alpha_\nu)
\end{equation}
and
\begin{equation}
\label{map_theta_iter}
\theta_t=\theta_0+\sum_{\mu=1}^t p_\mu
=\theta_0+p_0\,t+k\sum_{\nu=0}^{t-1}(t-\nu)\,\sin(\theta_\nu+\alpha_\nu)\ .
\end{equation}
Using the assumption of random and uncorrelated angles we find that 
the quantities $\delta p_t=p_t-p_0$ and 
$\delta \theta_t=\theta_t-\theta_0-p_0\,t$ are random gaussian variables 
(for $t\gg 1$) with average and variance:
\begin{equation}
\label{p_theta_average}
\langle \delta p_t\rangle=\langle \delta \theta_t\rangle=0\ ,
\end{equation}
\begin{equation}
\label{p_theta_variance}
\langle \delta p_t^2\rangle=\frac{k^2}{2}\,t\quad,\quad
\langle \delta \theta_t^2\rangle=\frac{k^2}{12} t(t+1)(2t+1)\approx 
\frac{k^2}{6}\,t^3\ ,
\end{equation}
implying a diffusive behavior in $p$-space with diffusion constant 
$D_0=(\Delta p)^2/\Delta t =k^2/2$ which 
is valid for $k\gg k_c$. For $k>k_c$ but close to $k_c$ we
expect the dynamics also to be diffusive but with a reduced diffusion constant 
$D<D_0$ due to correlations of $\theta_t$ for different times $t$ and for 
$k\le k_c$ we have $D=0$. However, we insist that this behavior is 
expected for long times scales, in particular $t\gg T$ with $T$ being 
the period of the random phases $\alpha_\nu=\alpha_{T+\nu}$. For small 
times $t\le T$ there is always, for arbitrary  values of $k$ 
(including the case $k<k_c$), a simple
short time diffusion with the diffusion constant $D_0$ and in this regime 
Eq. (\ref{p_theta_variance}) is actually exact (if the average is understood 
as the average with respect to $\alpha_\nu$).

It is interesting to note that Eq. (\ref{p_theta_variance}) provides an 
additional way to derive the chaos border $k_c$. Actually, we expect the long 
term dynamics to be diffusive if at the 
end of the first period $t=T$ the short time diffusion allows to cross 
at least one resonance in $p$ space~: 
$\delta p_T\sim k\sqrt{T}>\Delta p=2\pi/T$ or if the dynamics becomes 
ergodic in $\theta$ space~: 
$\delta \theta_T\sim k T^{3/2}>2\pi$. Both conditions provide the same 
chaos border $k>k_c\sim T^{-3/2}$ exactly confirming the finding of the 
previous section by the Chirikov criterion of overlapping resonances. 

We note that the behavior $\delta\theta_t\sim k\,t^{3/2}$ is a 
direct consequence that $\delta \theta_t$ is a sum (integral) over 
$\delta p_\mu$ 
for $\mu\le t$ and that $p_\mu$ itself is submitted to a diffusive dynamics 
with $\delta p_\mu \sim \mu^{1/2}$. The same type of phase fluctuations also 
happens in other situations, notably in quantum dynamics where the quantum 
phase is submitted to some kind of noise with diffusion in energy 
or in frequency space (with diffusion 
constant $D_0$) implying a dephasing time $t_\phi\sim D_0^{-1/3}$. 
For example, a similar situation appears for 
dephasing time in disordered conductors \cite{altshuler}
where the diffusive energy 
fluctuations for one-particle states are caused by electron-electron 
interactions and where  the same type of parametric dependence
(in terms of the energy diffusion constant) 
is known to hold. Another example is the 
adiabatic destruction of Anderson localization discussed 
in \cite{borgonovi} where a small noise in the hopping matrix 
elements of the one-dimensional Anderson model leads to a destruction of 
localization and diffusion in lattice space because the quantum 
phase coherence is limited by the same kind of mechanism.

\begin{figure}[ht]
\begin{center}
\includegraphics[width=0.48\textwidth]{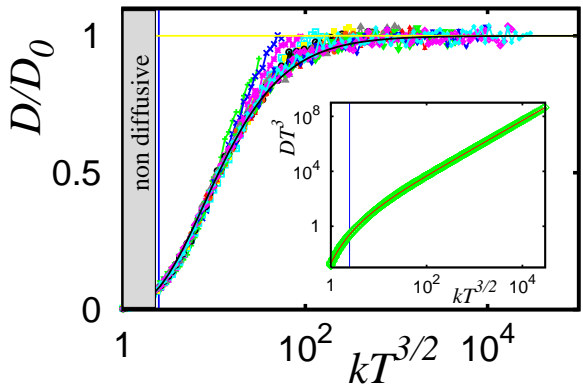} 
\end{center}
\caption{(Color online)
The ratio $D/D_0$ of the classical diffusion constant $D$ of 
the Chirikov typical map 
over the theoretical diffusion constant $D_0=k^2/2$, assuming perfectly 
uncorrelated phases $\theta_t$, as a function of the scaling parameter 
$x=kT^{3/2}$ for $10 \le T \le 1000$ and $T^{-3/2}\le k\le 1$. The 
classical diffusion constant $D$ has been obtained from a linear fit 
of the average variance $<\delta p_t^2>$ in the time interval 
$10\,T\le t\le 100\,T$. 
The average variance has been calculated for 400 different 
realizations of the random phase shifts and 25 different random 
initial conditions $(\theta_0,p_0)\in[0,2\pi[\times[0,2\pi[$ for each 
realization. The vertical full (blue) line represents the classical 
chaos border at $k_c T^{3/2}=\pi^2/4\approx 2.47$ (compare Fig. \ref{fig2})
and the grey rectangle at the left shadows the non-diffusive regime where 
the numerical procedure (incorrectly) yields small positive values of $D$ 
(see text).
The full curved (black) line represents the scaling 
function 
$D/D_0=f(x)$ as given by Eq. (\ref{scaling_function}). 
The inset shows $DT^3$ as a function of the scaling parameter $x=kT^{3/2}$ 
and the full (red) line represents the scaling function $x^2 f(x)/2$. The 
full vertical (blue) line again represents the classical chaos border.
\label{fig3}}
\end{figure}

We now turn to the discussion of our numerical study of the 
classical diffusion in the Chirikov 
typical map. We have numerically determined the classical diffusion 
constant for 
values of $10\le T\le 1000$ and $T^{-3/2}\le k\le 1$. For this we have 
simulated the classical map up to times $t\le 100\,T$ and calculated 
the long time diffusion constant $D$ as the slope from the 
linear fit of the variance $\langle \delta p_t^2\rangle$ 
for $10\,T\le t\le 100\,T$ 
(in order to exclude artificial effects due to the obvious short time 
diffusion with $D_0=k^2/2$). According to Fig. \ref{fig3} the 
ratio  $D/D_0$ can be quite well expressed as a scaling function of 
the quantity~:
$kT^{3/2}\sim k/k_c$:
\begin{equation}
\label{diff_scaling}
D\approx D_0 f\left(kT^{3/2}\right)
\end{equation}
where the scaling function $f(x)$ can be approximated by the fit :
\begin{equation}
\label{scaling_function}
f(x)=\exp\left(1-\sqrt{1+\frac{18.8}{x}+\frac{23.1}{x^2}}\right)\ .
\end{equation}
This fit has been obtained from a plot of $\ln(D/D_0)$ versus $y\equiv x^{-1}$ 
where the numerical data gives a linear behavior 
$\ln(D/D_0)\approx -A_1 y$ for $y\ll 1$ and 
$\ln(D/D_0)\approx -A_2 y$+const. for $y\gg 1$ with different 
slopes $A_1$ and $A_2$ which can 
be fitted by the ansatz $\ln(D/D_0)=1-\sqrt{1+C_1\,y+C_2\,y^2}$. The 
problem to obtain an analytical theory for this scaling function is 
highly non-trivial and subject to future research. However, we note 
that the numerical scaling function (\ref{scaling_function}) shows the 
correct behavior in the limits $x\to\infty$ and $x\to 0$. 
For $x=k\,T^{3/2} \ge 100$ we have $f(x)\approx 1$ implying $D\approx D_0$ for 
the strongly diffusive regime where phase 
correlations in $\theta_t$ can be neglected. 
For intermediate values $k_c T^{3/2}\approx 2.47<x<100$ there is a two scale 
diffusive regime with 
a short time diffusion constant $D_0$ for $t\le T$ and a reduced longer time 
diffusion constant $D<D_0$ for $t\gg T$. 

For $k<k_c$ ($x<2.47$) the long time diffusion constant is expected to 
be zero but the numerical fit procedure still results in small positive 
values (note that the scaling function vanishes very quickly 
in a non-analytical way $f(x)\sim \exp(-4.8/x)$ as $x\to 0$) 
simply because here the chosen fit interval $10\,T<t<100\,T$ is too small. In 
order to numerically identify the absence of diffusion 
it would be necessary to consider much longer 
iteration times. We also note that the variance $\langle \delta p_t^2\rangle$ 
shows a quite oscillatory behavior for $k<k_c$ and $t>T$ indicating 
the non-diffusive character of the dynamics despite the small positive 
slope which is obtained from a numerical fit (in a limited time interval). 
Therefore, we have shadowed in Fig. \ref{fig3} the regime $x<2.47$ 
by a grey rectangle in order to clarify that this regime is non-diffusive.

The important conclusion of Fig. \ref{fig3} is that it clearly confirms 
the transition to chaotic diffusive dynamics at values 
$k>k_c\sim T^{-3/2}$ and that in addition there is even an approximate 
scaling behavior in the parameter $x=kT^{3/2}\sim k/k_c$. 

\section{Lyapunov exponent and ergodic dephasing time scale}
\label{sec5}

We consider the chaotic regime $k_c<k<1$ where the diffusion rate is quite
slow and where we expect that the Lyapunov exponent is much smaller than unity 
implying that the exponential instability of the trajectories 
develops only after several iterations of the classical map. 
In order to study 
this in more detail, we rewrite the  map as~:
\begin{equation}
\label{classicalmap_generic}
p_{t+1}=p_t+f_t(\theta_t)\quad,\quad 
\theta_{t+1}=\theta_t+p_{t+1}
\end{equation}
where for the Chirikov typical map we have 
$f_t(\theta)=k\cos(\theta+\alpha_t)$ but in this section 
we would like to allow for more general periodic kick functions 
$f_t(\theta)$ with vanishing average (over $\theta$). In order 
to determine the Lyapunov exponent of this map we need to consider 
two trajectories $(\theta_t,p_t)$ and $(\tilde \theta_t,\tilde p_t)$ both 
being solutions of (\ref{classicalmap_generic}) with very close initial 
conditions at $t=0$. The differences $\Delta p_t=p_t-\tilde p_t$ and 
$\Delta\theta_t=\theta_t-\tilde\theta_t$ are iterated by the following 
linear map:
\begin{equation}
\label{linear_map}
\Delta p_{t+1}\approx \Delta p_t+f_t'(\theta_t)\Delta\theta_t
\quad,\quad 
\Delta \theta_{t+1}=\Delta \theta_t+\Delta p_{t+1}
\end{equation}
as long as $|\Delta\theta_t|\ll 1$. In a similar way as with the Eqs. 
(\ref{map_p_iter}) and (\ref{map_theta_iter}) in the last section, we may 
iterate the linear map up to times $t$~:
\begin{eqnarray}
\label{linear_p_iter}
\Delta p_t&=&\Delta p_0+\sum_{\nu=0}^{t-1}
f_\nu'(\theta_\nu)\Delta\theta_\nu\ ,\\
\nonumber
\Delta \theta_t&=&\Delta \theta_0+\sum_{\mu=1}^t \Delta p_\mu\\
\label{linear_theta_iter}
&=&\Delta \theta_0+t\,\Delta p_0+\sum_{\nu=0}^{t-1}(t-\nu)\,
f_\nu'(\theta_\nu)\,\Delta\theta_\nu\ . 
\end{eqnarray}
We now assume that in the chaotic regime the phases $\theta_\nu$ 
are random and uncorrelated, and that the average of the squared phase 
difference behaves as~:
\begin{equation}
\label{smooth_asumption}
\langle\Delta\theta_t^2\rangle=\Theta(t)\,\Delta\theta_0^2
\end{equation}
with a smooth function $\Theta(t)$ we want to determine. 
From Eq. (\ref{linear_theta_iter}) we obtain in the continuum limit 
the following integral equation for the function $\Theta(t)$~:
\begin{equation}
\label{integral_equaiont }
\Theta(t)=1+2A\,t+A^2\,t^2+\kappa\int_0^t(t-\nu)^2\Theta(\nu)\,d\nu
\end{equation}
where $\kappa=\langle f_\nu'(\theta)^2\rangle_\theta$ and 
$A=\Delta p_0/\Delta\theta_0$. 
For the map (\ref{classicalmap}) we have $\kappa=k^2/2$.
This integral equation implies the differential 
equation 
\begin{equation}
\label{Theta_differential_equation}
\Theta'''(t)=2\kappa\, \Theta(t)
\end{equation}
with the general solution~:
\begin{equation}
\label{Theta_general_solution}
\Theta(t)=\sum_{j=1}^3 C_j\,e^{2\lambda_j\,t}
\end{equation}
where the constants $C_j$ are determined by the initial conditions at $t=0$ 
and $2\lambda_j$ are the three solutions of $(2\lambda_j)^3=2\kappa$~:
\begin{equation}
\label{lambda_solution}
\lambda_1=(2\kappa)^{1/3}/2
\quad,\quad\lambda_2=e^{2\pi i/3}\lambda_1\quad,
\quad\lambda_3=\lambda_2^*\ .
\end{equation}
Since only $\lambda_1$ has a positive real part, we have in the long time 
limit~:
\begin{equation}
\label{Theta_limit_solution}
\Theta(t)\approx C_1\,e^{2\lambda_1\,t}
\end{equation}
and therefore by Eq. (\ref{smooth_asumption}) 
$\lambda_1$ represents approximately the Lyapunov exponent of the map 
(\ref{classicalmap_generic}). 

We note that this calculation of the Lyapunov exponent is not exact, 
essentially because we evaluate the direct average 
$\langle \Delta \theta_t^2\rangle$ instead of 
$\exp(\langle\ln(\Delta \theta_t^2)\rangle)$. 
A proper and more careful evaluation of the Lyapunov exponent for this 
kind of maps (in the regime $\kappa\ll 1$, assuming chaotic behavior 
with uncorrelated phases) has been done by 
Rechester et al. \cite{rechester} in the context of a motion along a 
stochastic magnetic field. Their result reads in our notations~:
\begin{equation}
\label{rechester_lyapunov}
\lambda=\frac{3^{1/3}\,\Gamma(\frac53)}{4\,\Gamma(\frac43)}\,\kappa^{1/3}
\approx 0.36\,\kappa^{1/3} \approx 0.29 k^{2/3}
\end{equation}
while from Eq. (\ref{lambda_solution}) we have 
$\lambda_1\approx 0.63\,\kappa^{1/3}$ with the same parametric dependence 
but with a different numerical prefactor. The dependence 
(\ref{rechester_lyapunov}) has been confirmed in numerical simulations
\cite{chi1981} for the model (\ref{classicalmap})
and we do not perform numerical simulations for $ \lambda$
here. We note that in the map (\ref{classicalmap})
the Lyapunov exponent 
$\lambda$ gives the Kolmogorov-Sinai entropy \cite{lichtenberg}.

The inverse of the Lyapunov exponent defines the Lyapunov time scale 
$t_{\rm Lyap}\sim \kappa^{-1/3}$ which is the time necessary to 
develop the exponential instability of the chaotic motion. 
According to Eq. (\ref{p_theta_variance}) we also have~:
$\langle \delta\theta_t^2\rangle=D_0\,t^3/3$ with 
$D_0=\langle f_\nu(\theta)^2\rangle_\theta$ implying an ergodic dephasing time 
$t_\Phi\sim D_0^{-1/3}$ being the time necessary for a complete dephasing 
where there is no correlation of the actual phase $\theta_t$ 
with respect to the 
ballistic phase $\theta_{\rm ball.}=\theta_0+p_0\,t$. 
For the Chirikov typical map with $f_\nu(\theta)=k\,\sin(\theta+\alpha_\nu)$ 
the averages of $f_\nu'(\theta)^2$ and $f_\nu(\theta)^2$ are identical, 
implying $\kappa=D_0=k^2/2$, and these two time scales coincide~: 
$t_{\rm Lyap}=t_\Phi\sim k^{-2/3}$. Furthermore, 
the condition for global chaos, 
$k>k_c\sim T^{-3/2}$, reads $t_{\rm Lyap}=t_\Phi<T$ implying 
that the exponential instability and complete dephasing must happen before 
the period $T$. 

We mention that for other type of maps, in particular if $f_\nu(\theta)$ 
contains higher harmonics such as $\ \sin(M\theta)\ $ we may have~: 
$\kappa \sim M^2 D_0$, and therefore parametrically different times scales 
$t_{\rm Lyap}\sim M^{-2}\,t_\Phi$. Examples of these type of maps have been 
studied in Ref. \cite{rechester} and also in Refs. \cite{frahm1,frahm1a} 
in the 
context of angular momentum diffusion and localization in rough billiards. 

For the Chirikov typical map there is a further time scale $t_{\rm Res}$ 
which is the time necessary to cross one resonance of width $\Delta p=2\pi/T$ 
by the diffusive motion $\langle\delta p_t^2\rangle=k^2\,t/2$~:
\begin{equation}
\label{t_resonance}
t_{\rm Res}\sim \frac{1}{(kT)^2}
\sim k^{-2/3}\,\left(\frac{k}{k_c}\right)^{-4/3} 
\sim t_\Phi\,\left(\frac{k}{k_c}\right)^{-4/3}\quad.
\end{equation}
In the chaotic regime $k>k_c$ this time scale is parametrically smaller 
than the dephasing time and the Lyapunov time scale.

\section{Quantum evolution}
\label{sec6}

We now study the quantum evolution which is described 
by the quantum map (\ref{quantummap}) (see section \ref{sec2}).
Typically in the literature studying the quantum version 
of Chirikov standard map 
the value of $\hbar$ is absorbed in 
a modification of the elementary kick period and the kick parameter $k$. 
Here we prefer to keep 
$\hbar$ as an independent parameter (also for the numerical simulations) 
and to keep the notation $T$ of the (integer) period of the time dependent 
kick-angle $\alpha(t)$. In this way we clearly 
identify two independent classical parameters $k$ and $T$, one quantum 
parameter $\hbar$ and a numerical parameter $N\gg 1$ being the 
finite dimension of the Hilbert space for the numerical quantum simulations.
For the physical understanding and discussion it is quite useful to well 
separate the different roles of these parameters and we furthermore avoid 
the need to translate between ``classical'' and ``quantum'' versions of the 
kick strength $k$.

We choose the Hilbert space dimension to be a power of 2~: $N=2^L$ 
allowing an efficient use of the discrete fast Fourier transform (FFT) in 
order to switch between momentum and position (phase) representation. 
A state $\ket{\psi}$ is represented by a complex vector 
with $N$ elements $\psi(j)$, $j=0,\ldots,N-1$ where $p_j=\hbar j$ are the 
discrete eigenvalues of the momentum operator $\hat p$. 
In order to apply the map (\ref{quantummap}) to the state $\ket{\psi}$ 
we first use an inverse FFT to transform to the phase representation 
in which the operator $\hat\theta$ is diagonal with eigenvalues 
$\theta_j=2\pi j/N$, $j=0,\ldots,N-1$, then we apply the first unitary matrix 
factor in (\ref{quantummap}) which is diagonal in this base, 
we apply an FFT to go back to the momentum 
representation and finally the second unitary matrix factor, 
diagonal in the momentum representation, is applied. 
This procedure can be done with ${\cal O}(N\log_2(N))$ operations 
(for the FFT and its inverse) plus ${\cal O}(N)$ operations for the 
application of the diagonal unitary operators. 

We have also to choose a numerical value of $\hbar$ and this depends on which 
kind of regime (semiclassical regime or strong quantum regime) we want to 
investigate. We note that due to the dimensional 
cut off we obtain a quantum periodic boundary condition~: $\psi(0)=\psi(N)$, 
i.~e. the quantum dynamics provides in $p$ representation always 
a momentum period of $\hbar N$. Furthermore the classical map is also 
periodic in momentum with period $2\pi$ and it is possible to 
restrict the classical momentum to one elementary cell where the momentum 
is taken modulo $2\pi$. Therefore one 
plausible choice of $\hbar$ amounts to choose the 
quantum period $\hbar N$ to be equal to the classical period $2\pi$, i.~e.: 
$\hbar =2\pi/N$. In this case the quantum state covers exactly one elementary 
cell in phase space. Since $\hbar\to 0$ for $N\to\infty$, we refer to this 
choice as the semiclassical value of $\hbar$. 

\begin{figure}[ht]
\begin{center}
\includegraphics[width=0.48\textwidth]{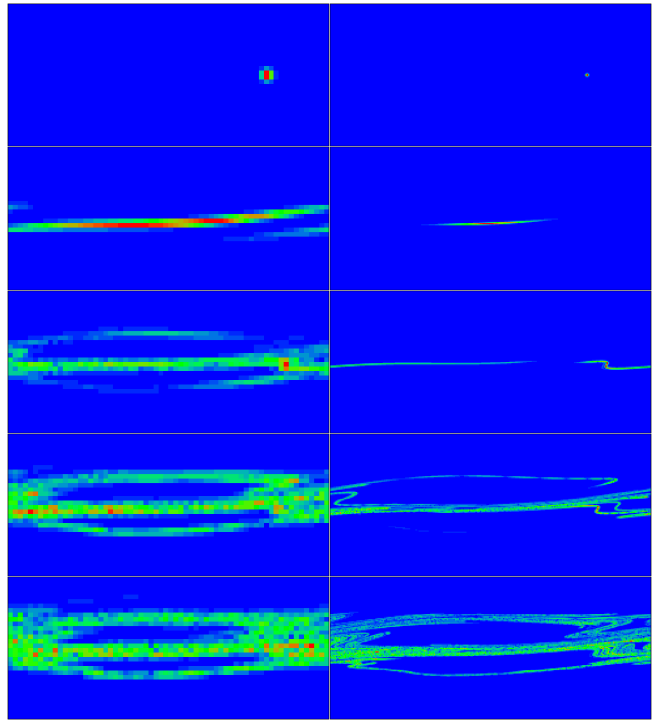} 
\end{center}
\caption{(Color online)
Quantum evolution of the Chirikov typical map at $k=0.1$ and $T=10$ 
in the semiclassical limit 
with $\hbar=2\pi/N$ and with the Hilbert space dimension 
$N=2^{12}$ (left column) and $N=2^{16}$ (right column) 
at times $t=0$ (first row), $t=20$ (second row), $t=60$ (third row), $t=100$ 
(fourth row) and $t=150$ (fifth row). Shown are Husimi functions with maximum 
values at red (grey), intermediate values at green
(light grey) and minimum values at blue (black)
at the lower half of one elementary cell $\theta\in [0,2\pi[$, 
$p\in [0,\pi[$. The 
initial condition is a coherent gaussian state centered at 
$\theta_0=0.8\cdot 2\pi$ and $p_0=0.25\cdot 2\pi$ with a variance (in 
$p$-representation) of $\Delta p=2\pi/\sqrt{12N}$. The resolution corresponds 
to $\sqrt{N}$ (64 or 256) squares in one line. The realization of the 
random phase shifts is identical to that of Fig. \ref{fig2}. 
Note that at the considered value $k=0.1$ the global classical 
dynamic is diffusive but 
requires iteration times of $t\sim 10000$ to fill one elementary cell
(see Fig. \ref{fig5}).
\label{fig4}}
\end{figure}

In this section, we present some numerical results of the quantum dynamics 
using the semiclassical value of $\hbar$. In Fig.~\ref{fig4}, we show 
the Husimi functions for the case $T=10$, $k=0.1$ with 
an initial state being a minimal gaussian wave packet centered at 
$\theta_0=0.8\cdot 2\pi$ and $p_0=0.25\cdot 2\pi$ with a variance (in 
$p$-representation) of $\Delta p=2\pi/\sqrt{12N}$. This position 
is well inside a classically chaotic region (see Fig. \ref{fig2}). 
The Hilbert space dimensions $N$ are $2^{12}$ and $2^{16}$ and the 
iteration times are $t\in\,\{0,\,20,\,60,\,100,\,150\}$. 
We see that at these time scales the motion extends to two 
classical resonances with two stable and quite large islands. For 
$N=2^{12}$ the classical phase space structure is quite visible but 
the finite ``resolution'' in phase space due to quantum effects is quite 
strong while for $N=2^{16}$ the Husimi function allows to resolve 
much better smaller details of the classical motion. 
We note that the Husimi function is obtained 
by smoothing of the Wigner function
over a phase space cell of $\hbar$ size (see e.g. more detailed
definitions  and Refs. in \cite{frahm2}).

\begin{figure}[ht]
\begin{center}
\includegraphics[width=0.48\textwidth]{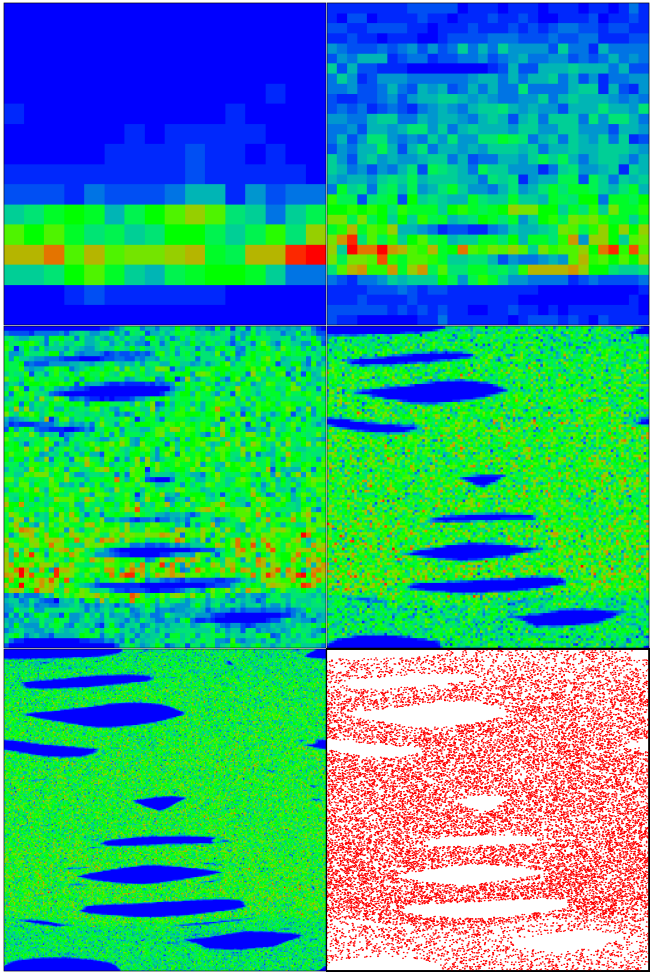} 
\end{center}
\caption{(Color online)
Quantum evolution of the Chirikov typical map as in Fig. \ref{fig4} 
with the same 
coherent gaussian state as 
initial condition, same values $k=0.1$ and $T=10$, $\hbar=2\pi/N$ but 
at the iteration time $t=20000$ and at different values of $N=2^8$, 
$N=2^{10}$ (first row), $N=2^{12}$, $N=2^{14}$ (second row),
$N=2^{16}$ and classical simulation (third row). For the classical map 
20000 trajectories have been iterated up to the same time $t=20000$ 
with random initials conditions very close to the initial position at 
$\theta_0=0.8\cdot 2\pi \pm 0.002$ and 
$p_0=0.25\cdot 2\pi \pm 0.002$. Colors are as in Fig.~\ref{fig4}.
\label{fig5}}
\end{figure}

We note that for $T=10$ the value of the kick strength 
$k=0.1$ is only slightly above the chaos border $0.078$ 
with global diffusion but there are still large stable islands 
that occupy a significant fraction of the phase space. 
At this value the numerically computed
diffusion constant is $D=0.1 D_0\approx 1/2000$ 
(see section \ref{sec3} and Fig. \ref{fig3}). In 
Fig. \ref{fig5} we compare the Husimi functions for various Hilbert space 
dimensions at time $t=20000$. This time is 
sufficiently long so that  a diffusive spreading 
in $p$-space gives $\delta p_t=\sqrt{Dt}\approx 3.162$ 
thus roughly covering one elementary cell (but not absolutely uniformly). 

For the smallest value of $N=2^8$ we observe a very strong influence of 
quantum effects with a Husimi function extended to half 
an elementary cell which is significantly stronger localized 
(in the momentum direction) than the 
diffusive classical spreading would suggest. Furthermore 
we cannot identify any classical phase space structure. 
This is quite normal due to the very limited resolution of $\sqrt{N}=16$ 
``quantum-pixels'' in both directions of $\theta$ and $p$.
For $N=2^{10}$ the quantum effects are still strong but the Husimi function 
already extends to $~85$\% of the elementary cell and we can identify 
first very slight traces of at least three large stable islands.
For $N=2^{12}$ the Husimi function fills the elementary cell as suggested 
by the classical spreading but the distribution is less uniform than for 
larger values of $N$ or for the classical case. We can also 
quite well identify the large scale 
structure of the phase space with the main stable islands associated 
to each resonance. However, the fine structure of phase space is not visible 
due to quantum effects. 
For $N=2^{14}$ and even more for $N=2^{16}$, the resolution of the Husimi 
function increases and approaches the classical distribution which is 
also shown in Fig. \ref{fig5} for comparison. For $N=2^{16}$, we even 
see first traces of small secondary islands that are not associated 
to the main resonances. 
We note that the classical distribution in Fig. \ref{fig5} 
is not a full phase portrait (showing ``all'' 
iteration times) but it only contains the classical positions 
after $t=20000$ iterations with random initial positions~: 
$\theta_0=0.8\cdot 2\pi\pm 0.002$, $p_0=0.25\cdot 2\pi\pm 0.002$
close to the gaussian wave packet used as initial state 
in Figs. \ref{fig4}, \ref{fig5}. 

Finally we note that a wave packet with initial size 
$\delta \theta_0 \approx \sqrt{\hbar}$ 
grows exponentially with time
and spreads over the whole phase interval $2\pi$
after the Ehrenfest time scale \cite{chi1981,zasl,chi1988}
\begin{equation}
\label{te}
t_{\rm E} \approx \ln(2\pi/\sqrt{\hbar})/\lambda .
\end{equation}
For the parameters of Fig.~\ref{fig4}, e.g. $\hbar=2\pi/2^{16}$,
this gives $t_{\rm E} \approx 103  $ that is in agreement with the
numerical data showing that the spearing in phase
reaches $2\pi$ approximately at $t=100$.

In summary the quantum simulation of the Chirikov typical map 
using the semiclassical value 
$\hbar=2\pi/N$ reproduces quite well the classical phase space structure 
for sufficiently large $N$ while for smaller values of the Hilbert space 
dimension ($N\le 2^8$) the nature of motion remains strongly quantum
and the diffusive spreading over the cell is stopped by quantum localization.

\section{Chirikov Localization}
\label{sec7}

It is well established \cite{chidls,dls1986} that, in general, quantum maps 
on one-dimensional lattices, whose classical counterpart is diffusive 
with diffusion constant 
$D_{\rm cl}$, show dynamical exponential localization of the eigenstates 
of the unitary map operator with the localization length 
$\ell_0=D_{\rm cl.}/\hbar^2$ 
measured in number of lattice sites 
(that corresponds to the quantum number $n$
associated to the momentum by $p=\hbar n$). 
Here $D_{\rm cl.}$ is the diffusion rate in action
per  period of the map.
This expression for $\ell_0$ 
is valid for the unitary symmetry class which applies to the Chirikov 
typical map which is not symmetric with respect to the transformation 
$\theta\to -\theta$. We have furthermore 
to take into account that the map (\ref{quantummap}) depends on time due 
to the random phases $\alpha_\nu$ and in order to determine the localization 
length we have to use the diffusion constant for the full $T$-times 
iterated map (which does not depend on time)~: $D_{\rm cl.}=D_0\,T$ assuming 
we are in the regime where $D=D_0$ for $k\gg k_c$. In this case we expect a 
localization length 
\begin{equation}
\label{loc_expect}
\ell_0=\frac{D_0\,T}{\hbar^2}=\frac{k^2\,T}{2\hbar^2}\ .
\end{equation}
This 
localization length is obtained as the exponential localization length 
from the eigenvectors of the full ($T$-times iterated) unitary map operator.
In numerical studies of the Chirikov typical map
it is very difficult and costly to access to these eigenvectors and 
we prefer to simply iterate the quantum map with an initial state localized 
at one momentum value in $p$-representation and to measure the 
exponential spreading of 
$\ket{\psi(t)}$ at sufficiently large times (using time and ensemble 
average). This procedure is known \cite{dls1986} 
to provide a localization length $\ell$ artificially enhanced by a 
factor of two: $\ell=2\ell_0$. 

Let us note that the relation (\ref{loc_expect})
assumes that the classical dynamics is chaotic and is characterized
by the diffusion $ D_0$. It also assumes that the eigenvalues
of the unitary evolution operator are homogeneously
distributed on the unitary circle. For certain dynamical 
chaotic systems the second condition can be violated giving rise
to a multifractal spectrum and delocalized eigenstates
(this is e.g. the case of the kicked Harper model
\cite{lima,prosen}). The analytical derivation of  (\ref{loc_expect})
using supersymmetry field theory assumes directly \cite{frahm1a}
or indirectly \cite{zirnb} that
the above second condition is satisfied. We also assume 
that this condition is 
satisfied due to randomness of $\alpha_t$ 
phases in the map (\ref{classicalmap}).

Before we discuss our numerical results for the localization length 
we would like to remind the 
phenomenological argument coined in \cite{chi1981} which allows 
to determine the above expression relating 
localization length to the diffusion constant and which we will below 
refine in order to take into account the two-scale diffusion with different 
diffusion constants at short and long time scales.

Suppose that the classical spreading in $p$-space is given by a known 
function~:
\begin{equation}
\label{spreading_function}
P_2(t)=\langle \delta p_t^2\rangle
\end{equation}
where $P_2(t)$ is a linear function for simple diffusion but it may be more 
general in the context of this argumentation (but still below the ballistic 
behavior $t^2$ for $t\to\infty$). For the quantum dynamics 
we choose an initial state localized at one momentum value. This state 
can be expanded using $\ell$ eigenstates of the full map operator 
with a typical eigenphase spacing $2\pi/\ell$. The iteration time $t$ 
(with $t$ being an integer multiple of $T$) corresponds to 
$t/T$ applications of the full map operator. We expect the quantum dynamics to 
follow the classical spreading law (\ref{spreading_function}) 
for short time scales such that we cannot 
resolve individual eigenstates of the full map operator, i.~e. for 
$2\pi t/(T\ell)<2\pi$. Therefore at the critical time scale 
$t^*=\ell T$ we expect the effect of quantum localization to set in 
and to saturate the classical spreading at the value 
$\delta p_{\rm loc.}^2\sim\hbar^2\ell^2$ due to the finite 
localization length $\ell$ (measured in integer units of 
momentum quantum numbers). 
This provides an implicit equation for $t^*$~:
\begin{equation}
\label{implicit_tstar}
\left(\frac{t^*}{T}\right)^2=\ell^2=C\frac{P_2(t^*)}{\hbar^2}
\end{equation}
where $C$ is a numerical constant of order unity. 
Let us first consider the case of simple diffusion with constant $D_0$ 
for which we have $P_2(t)=D_0\,t$. In this case we obtain~:
\begin{equation}
\label{loc_simplediffusion}
\ell=\frac{t^*}{T}=C\frac{D_0\,T}{\hbar^2}=C\ell_0
\end{equation}
with $\ell_0$ given by Eq. (\ref{loc_expect}). This result provides 
the numerically measured value $\ell=2\ell_0$ by the exponential 
spreading of $\ket{\psi(t)}$ if we choose $C=2$. Below we will also apply 
Eq. (\ref{implicit_tstar}) to the case of two scale diffusion with 
different diffusion constants at short and long time scales providing 
a modified expression for the localization length. 

\begin{figure}[ht]
\begin{center}
\includegraphics[width=0.48\textwidth]{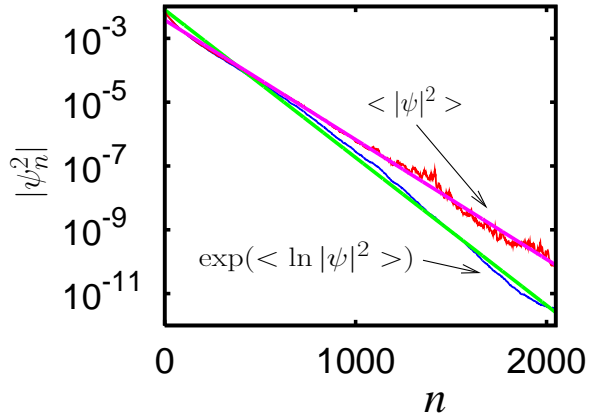} 
\end{center}
\caption{(Color online)
Illustration of the Chirikov localization for the quantum Chirikov 
typical map. Shown 
is the (averaged) absolute squared wave function in $p$-representation 
as a function of the integer quantum number $n$ associated to $p=\hbar n$. 
The initial state at $t=0$ is localized at $n=0$ and the quantum typical 
map has been applied up to times $t=t_{\rm max}$ with $t_{\rm max}$ chosen 
such that the initial classical diffusion is well saturated 
at $t_{\rm max}/4$ due to quantum effects 
(see Fig. \ref{fig7}). In order to determine numerically the localization 
length, first $|\psi_n|^2$ has been time-averaged for the time interval 
$t_{\rm max}/4\le t\le t_{\rm max}$ and then a further ensemble average (over 
100 realizations of the random phase shifts) of $|\psi_n|^2$ 
(upper red/gray curve) 
or of $\ln(|\psi_n|^2)$ (lower blue/black curve) has been applied. 
The localization
lengths are obtained as the (double) inverse slopes of a linear fit 
(with relative weights $w_n\sim |\psi_n|^2$) of $\ln(|\psi_n|^2)$ versus $n$. 
Here the parameters are: $N=2^{12}$, $k=0.5$, $T=100$, 
$t_{\rm max}=546132$, 
$\hbar=2\pi/(17+\gamma)\approx 0.357$ with the golden number 
$\gamma=(\sqrt5-1)/2\approx 0.618$. 
The fits are given by the thick straight lines. 
The theoretical localization length is 
$\ell_0=k^2T/(2\hbar^2)\approx 98.3$ while the numerical fits provide 
$\ell_\psi\approx 231.7$ (upper curve) and $\ell_{\ln\psi}\approx 187.5$
(lower curve). 
\label{fig6}}
\end{figure}

First we want to present our numerical results for the localization length 
of the quantum Chirikov typical map. Since the localization length scales 
as $\hbar^{-2}$ we cannot use the semiclassical value $\hbar=2\pi/N$ 
since in the limit $N\to\infty$ the localization length would always be 
larger than $N$ and in addition we would only cover one 
elementary classical cell 
of phase space which is not very suitable to study the effects of diffusion 
and localization in momentum space. Therefore we choose a finite and 
fixed value $\hbar=2\pi/(\tilde N+\gamma)$ where $\tilde N$ is 
some fixed integer in the range $1\ll \tilde N\ll N$ and 
$\gamma=(\sqrt5-1)/2\approx 0.618$ is the golden number because we want 
to avoid artificial resonance effects between the classical momentum 
period $2\pi$ and the quantum period $\hbar N$ (due to the finite dimensional 
Hilbert space). The ratio of these two periods, $N/(\tilde N+\gamma)\gg 1$, 
is roughly the number of elementary classical cells covered by the quantum 
simulation. For most simulations we have chosen $\tilde N=17$ 
and varied the classical parameters $k$ and $T$ but we also provide the 
data for a case where the values of $k$ and $T$ are fixed and 
$\tilde N$ varies. 

For the numerical quantum simulation we choose the initial state 
$\ket{\psi(0)}=\ket{0}$ being perfectly localized in momentum space and we 
apply the quantum Chirikov typical map up to a sufficiently large time scale 
$t_{\rm max}$ chosen such that the initial diffusion is well 
saturated at $t_{\rm max}/4$ and we perform a time average 
of $|\psi_n|^2=|\bra{n}\ket{\psi}|^2$ for the 
time interval $t_{\rm max}/4\le t\le t_{\rm max}$. The resulting 
time average is furthermore averaged with respect to different realizations 
of the $T$ random phases $\alpha_\nu$ and here we consider two cases where 
we average either $|\psi_n|^2$ or $\ln(|\psi_n|^2)$. We then 
determine two numerical values $\ell_\psi$ and $\ell_{\ln\psi}$ by a 
linear fit of $\ln(|\psi_n|^2)=2n/\ell+$const. with relative weight factors 
$w_n\sim|\psi_n|^2$ in order to emphasize the initial exponential decay 
and to avoid problems at large values of $n$ where the finite numerical 
precision ($\sim 10^{-15}$) or the finite value of $N$ may create 
an artificial saturation of the exponential decay. This procedure is 
illustrated in Fig. \ref{fig6} for the parameters $k=0.5$, $T=100$, 
$\hbar=2\pi/(17+\gamma)$ and $N=2^{12}$. We see that the two numerical values 
$\ell_\psi$ and $\ell_{\ln\psi}$ are roughly $2\ell_0$ as expected 
but may differ among themselves by a modest numerical factor. 

\begin{figure}[ht]
\begin{center}
\includegraphics[width=0.48\textwidth]{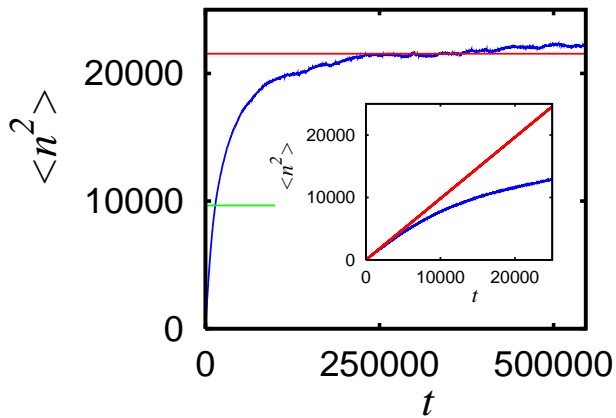} 
\end{center}
\caption{(Color online)
The ensemble averaged variance of the expectation value of $n^2$ as a function 
of $t$ for the same parameters (and the same realizations of the random phase 
shifts) as in Fig. \ref{fig6}. The upper horizontal 
(red) line corresponds to the saturation value $\ell_{\rm diff}^2$ (obtained 
from a time average of $<n^2>$ in the interval 
$t_{\rm max}/4\le t\le t_{\rm max}$) with 
$\ell_{\rm diff}\approx 146.8$ being the localization due to 
{\em saturation of diffusion}. The short lower horizontal (green) line
corresponds to $\ell_0^2$ with the theoretical localization length 
$\ell_0\approx 98.3$ (see also Fig. \ref{fig6}).
The inset shows the initial diffusive regime at shorter time scales. The 
straight (red/grey) line shows the value $D_0/\hbar^2\approx 0.983$ 
associated to the initial diffusion: $\hbar^2\langle n^2\rangle=
\langle p^2\rangle\approx D_0\,t$ and 
with the theoretical diffusion constant $D_0=k^2/2$.
\label{fig7}}
\end{figure}

The localization length can also be determined from the saturation 
of the diffusive spreading which gives a localization length 
$\ell_{\rm diff}$ where $\ell_{\rm diff}^2$ is the time average 
of the quantum expectation value $\langle n^2\rangle$ for the interval 
$t_{\rm max}/4\le t\le t_{\rm max}$ (see Fig. \ref{fig7}). Typically 
$\ell_{\rm diff}$ is comparable to $\ell_\psi$ and $\ell_{\ln\psi}$ up to 
a modest numerical factor.

\begin{figure}[ht]
\begin{center}
\includegraphics[width=0.48\textwidth]{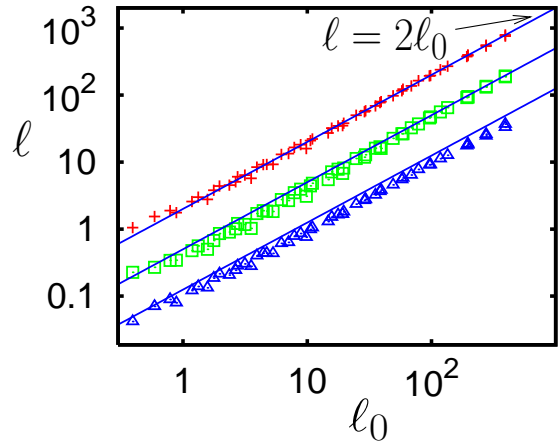} 
\end{center}
\caption{(Color online)
Localization length determined by three different numerical methods as a 
function of the theoretical value $\ell_0=k^2T/(2\hbar^2)$ in 
a double logarithmic representation. The data points correspond to 
$\ell_\psi$ (red crosses) obtained from a linear fit of 
$\ln\langle |\psi_n|^2\rangle$ versus 
$n$ (see Fig. \ref{fig6}), $\ell_{\ln\psi}$ (green squares) obtained 
from a linear fit of $\langle \ln|\psi_n|^2\rangle$ versus $n$, 
$\ell_{\rm diff}$ (blue triangles) obtained from the saturation of 
quantum diffusion (see Fig. \ref{fig7}). The straight full (blue) line 
represents $\ell=2\ell_0$. For clarity the data points for 
$\ell_{\ln\psi}$ 
(or $\ell_{\rm diff}$) have been shifted down by a factor of 4 (or 16).
The value $\hbar=2\pi/(17+\gamma)\approx 0.357$ is exactly as in Fig. 
\ref{fig6} while $k\in\{0.1, 0.15, 0.2, 0.3, 0.5, 0.7, 1.0\}$ and 
$T\in\{10, 15, 20, 30, 50, 70, 100\}$. The Hilbert space dimension 
is mostly $N=2^{12}$ except for a few data points with the largest values 
of $\ell_0$ where we have chosen $N=2^{14}$. The ensemble averages have been 
performed over 100 different realizations of the random phase shifts.
\label{fig8}}
\end{figure}

In Fig. \ref{fig8}, we compare the three numerically calculated 
values of the localization $\ell_\psi$, 
$\ell_{\ln\psi}$ and $\ell_{\rm diff}$ with the theoretical 
expression $\ell_0=D_0\,T/\hbar^2$ for various values of the 
classical parameters~: $k\in\{0.1, 0.15, 0.2, 0.3, 0.5, 0.7, 1.0\}$ and 
$T\in\{10, 15, 20, 30, 50, 70, 100\}$ and for the fixed value 
$\hbar=2\pi/(17+\gamma)\approx 0.357$. We see that $\ell_\psi$ 
agrees actually very well with $2\ell_0$ for a wide range of 
parameters and three orders of magnitude
variation. For $\ell_{\ln\psi}$ and 
$\ell_{\rm diff}$ the values are somewhat below $2\ell_0$ with a modest 
numerical factor (about 1.5 or smaller) but the overall dependence 
on the parameters is still correct on all scales. 

\begin{figure}[ht]
\begin{center}
\includegraphics[width=0.48\textwidth]{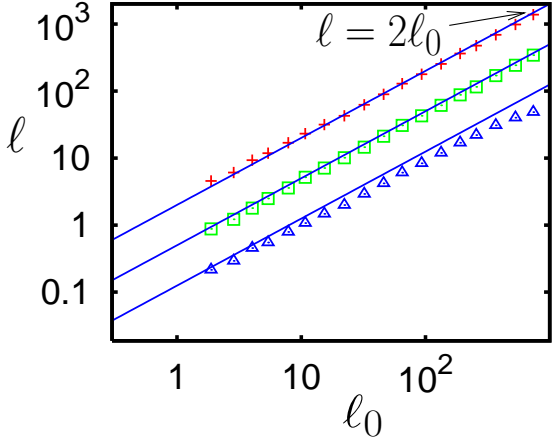} 
\end{center}
\caption{(Color online)
Same as in Fig. \ref{fig8} but with fixed values for the classical 
parameters $k=0.2$ and $T=50$ while $\hbar=2\pi/(\tilde N+\gamma)$ 
varies with the integer variable $\tilde N$ in the interval 
$8\le \tilde N\le 200$ 
(i.~e.: $0.03132\le \hbar\le 0.7291$). The number of realizations of the 
random phase shifts is 20.
\label{fig9}}
\end{figure}

In Fig. \ref{fig9}, we present a similar comparison as in Fig. \ref{fig8}, 
but here we have fixed the classical parameters to $k=0.2$ and 
$T=50$ and we vary $\hbar=2\pi/(\tilde N+\gamma)$ 
with $8\le \tilde N \le 200$ i.~e.: $0.03132\le \hbar\le 0.7291$. 
Again there is very good agreement of $\ell_\psi$ and 
$\ell_{\ln\psi}$ with $2\ell_0$ for nearly three orders of magnitude 
while for $\ell_{\rm diff}$ there is slight decrease for larger 
values of $\ell_0$. 

\begin{figure}[ht]
\begin{center}
\includegraphics[width=0.48\textwidth]{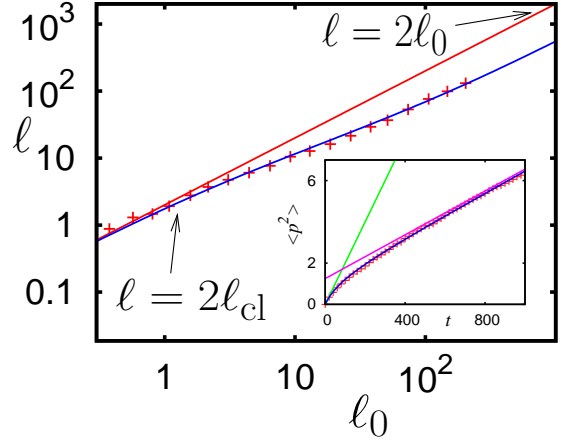} 
\end{center}
\caption{(Color online)
The localization length $\ell=\ell_\psi$ versus the theoretical value 
$\ell_0=k^2T/(2\hbar^2)$ for $k=0.2$, $T=10$ and the same values of 
$\hbar$ as in Fig. \ref{fig9}. The straight full (red) upper line corresponds 
to $\ell=2\ell_0$. The lower full (blue) curve corresponds to the corrected 
expression $\ell=2\ell_{\rm cl}(\ell_0)=2\ell_0/
\left[h(\ell_0)+\sqrt{h(\ell_0)^2+2B\ell_0}\right]$, 
$h(\ell_0)=(1-2A\ell_0)/2$, $A\approx 0.03121$, and 
$B\approx 0.1177$ due to a modified classical dynamics (see text for 
explanation).
The number of random phase realizations is 50. 
The inset shows the {\em classical diffusion} for the same classical 
parameters $k=0.2$, $T=10$. The data points (red crosses) are (selected) 
numerical data and the full (blue/black) curve represents the numerical 
fit: $\langle p^2\rangle=D_0\,t\,(T+At)/(T+Bt)$ providing the two parameters 
$A$ and $B$ for the corrected localization length $\ell_{\rm cl}$ in the main 
figure. The two straight lines correspond to the 
initial diffusion at short time scales with diffusion constant 
$D_0=k^2/2$ and the final diffusion at long time 
scales with $D=D_0\,(A/B)\approx 0.2652\,D_0$. This value of $D$ coincides 
quite well with the value of the scaling curve in Fig. \ref{fig3} at 
$x=kT^{3/2}\approx 6.32$. 
\label{fig10}}
\end{figure}

The agreement $\ell_\psi\approx 2\ell_0=2D_0\,T/\hbar^2$ in Fig. \ref{fig8} 
for a very large set of classical parameters is actually 
{\em too perfect} because some of the data points fall in the regime 
where the scaling parameter $x=kT^{3/2}$ is relatively small 
(between 2.47 and 30) with a classical diffusion constant $D$ well below 
its theoretical value $D_0=k^2/2$ (see Fig. \ref{fig3}). Therefore one 
should expect that the localization length is reduced as well according 
to $\ell\approx 2D\,T/\hbar^2<2\ell_0$ but the numerical data in 
Fig. \ref{fig8} do not at all confirm this reduction of the localization 
length we would expect from a classically reduced diffusion constant. 
One possible explanation is that the classical mechanism of 
relatively strongly correlated phases which induces the reduction 
of the diffusion constant depends on the fine structure of the classical 
dynamics of the Chirikov typical map in phase space, a fine structure which 
the quantum dynamics may not 
resolve if the value of $\hbar$ is not sufficiently low. Therefore 
the classical phase correlations are destroyed in the quantum simulation 
and we indeed observe the localization length $\ell=2\ell_0$ using the 
{\em theoretical} value of the diffusion constant $D_0$ and not the reduced 
diffusion constant $D$.
 
In order to investigate this point more thoroughly one should therefore vary 
$\hbar$ in order to see if it is possible to see this reduction 
of the diffusion constant also in the localization length 
provided that $\hbar$ 
is small enough. In Fig. \ref{fig9}, we have indeed data points with smaller 
values of $\hbar$ but here the classical parameters $k=0.2$ and $T=50$ still 
provide a large scaling parameter $k\,T^{3/2}\approx 70.7$ with a diffusion 
constant $D\approx 0.9\,D_0$ already quite close to $D_0$. 
In Fig. \ref{fig10}, 
we therefore study the same values of $\hbar$ (as in Fig. \ref{fig9}) 
but with modified classical parameters $k=0.2$ and $T=10$ such that 
$k\,T^{3/2}\approx 6.32$ resulting in a diffusion constant 
$D=0.2859\,D_0$ well below $D_0$. In Fig. \ref{fig10}, we indeed 
observe that the localization length $\ell_\psi$ is significantly below 
$2\ell_0$ for larger values of $\ell_0$ (small values of $\hbar$). 

We can actually refine the theoretical expression of the localization 
length in order 
to take into account the reduction of the diffusion constant. 
For this, we remind 
that for short time scales $t\le T$ the initial diffusion is always with 
$D_0$ and that only for $t\gg T$ we observe the reduced diffusion constant 
$D$. We have therefore applied the following fit~:
\begin{equation}
\label{spreading_fit}
P_2(t)=D_0\,t\,\frac{T+A\,t}{T+B\,t}
\end{equation}
to the classical spreading where $A$ and $B$ are two fit parameters. 
This expression fits actually very well the classical two scale diffusion 
with $D_0$ for short time diffusion and with $D=D_0(A/B)$ for the long 
time diffusion. For $k=0.2$ and $T=10$ we obtain (see inset of Fig. 
\ref{fig10}) the values $A\approx 0.03121$ and 
$B\approx 0.1177$ implying $D=D_0\,(A/B)\approx 0.2652\,D_0$ which is only 
slightly below the above value $D=0.2859\,D_0$ (obtained from the linear 
fit of the classical spreading for the interval $10\,T<t<100\,T$). 
We can now determine a refined expression of the localization length 
using the two scale diffusion fit (\ref{spreading_fit}) together 
with the implicit equation (\ref{implicit_tstar}) for the critical 
time scale $t^*$ which results in the following equation for $\ell$~:
\begin{equation}
\label{implicit_refine}
\ell^2=2\frac{P_2(\ell T)}{\hbar^2}=2\,\ell_0\,\ell
\frac{1+A\ell}{1+B\ell}\ .
\end{equation}
This is simply a quadratic equation in $\ell$ whose positive solution 
can be written in the form~:
\begin{equation}
\label{ell_refine}
\ell=2\ell_{cl}(\ell_0)=
\frac{2\ell_0}{h(\ell_0)+\sqrt{h(\ell_0)^2+2B\ell_0}}
\end{equation}
with $h(\ell_0)=(1-2\,A\ell_0)/2$. One easily verifies that the 
limit $D=D_0$, which corresponds to $A=B$, immediately reproduces 
$\ell_{\rm cl}(\ell_0)=\ell_0$ as it should be. Furthermore, the 
limit $\ell_0 A\gg 1$ (i.e.~: $\hbar\ll \sqrt{D_0\,T/A}$) provides 
$\ell_{\rm cl}(\ell_0)\approx \ell_0(A/B)$ while for $\ell_0 A\ll 1$ 
we have $\ell_{\rm cl}(\ell_0)\approx \ell_0$ (even for $A\ne B$).

The data points 
of $\ell_\psi$ in Fig. \ref{fig10} coincide very well with the 
refined expression (\ref{ell_refine}) thus clearly confirming the influence 
of the classical two scale diffusion on the value of the localization length
as described by Eq. (\ref{ell_refine}). 
Depending on the values of $\hbar$ the refined localization length 
$\ell_{\rm cl}(\ell)$ is either 
given by $\ell_0$ if $\hbar\gg \sqrt{D_0\,T/A}$ or by the reduced value 
$\ell_0(A/B)$ if $\hbar\ll \sqrt{D_0\,T/A}$. 
In the first case we do not see the effect of the reduced diffusion 
constant because 
the value of $\hbar$ is too large to resolve the subtle fine structure 
of the classical dynamics. Furthermore the critical 
time scale $t^*$, where the localization sets in, is below $T$ and the 
momentum spreading saturates already in the regime of the initial short time 
diffusion with $D_0$. 
In the second case $\hbar$ is small enough to resolve the fine structure of 
the classical dynamics and the time scale $t^*$ is 
above $T$ such that we may see the reduced diffusion constant $D=D_0(A/B)$ 
leading to the 
reduced localization length $\ell_{\rm cl}(\ell_0)=\ell_0(A/B)$. 

\section{Discussion}
\label{sec8}
In this work we analyzed the properties of classical and quantum Chirikov typical
map. This map is well suited to describe systems with continuous
chaotic flow. For the classical dynamics our studies established the dependence of diffusion 
and instability on system parameters being generally in 
agreement with the first studies presented in
\cite{chirikov1,rechester,chi1981}. In the quantum case we showed that the chaotic diffusion 
is localized by quantum interference effects giving rise to the Chirikov localization
of quantum chaos. We demonstrated that the localization length
is determined by the diffusion rate in agreement with the general
theory of Chirikov localization developed in \cite{chi1981,prange,chidls,dls1986}.
The Chirikov typical map has more rich properties compared to the Chirikov standard
map and we think that it will find interesting
applications in future.


\end{document}